\documentclass{appolb}
\usepackage{graphicx}

\begin{document}
\title{Development of J-PEM for breast cancer detection and diagnosis using positronium imaging}
\author{Shivani${}^1$\thanks{shivani.shivani@doctoral.uj.edu.pl}, El\.zbieta {\L}uczy\'nska${}^2$, Sylwia Heinze${}^3$, Pawe{\l} Moskal${}^1$\\ 
{\it\normalsize ${}^1$Faculty of physics and Applied Computer Science}\\ 
{\it\normalsize Jagiellonian University, Cracow,Poland }\\ 
{\it\normalsize ${}^2$Medical College of Rzeszow University}\\ 
{\it\normalsize Cracow, Poland}\\ 
{\it\normalsize ${}^3$Department of Radiology, Center of Oncology}\\ 
{\it\normalsize Cracow Branch, Cracow, Poland}} 

\maketitle
\begin{abstract}
The purpose of the presented investigations is to design, construct and establish the characteristic performance of the Jagiellonian Positron Emission Mammography(J-PEM), being designed for the detection and diagnosis of breast cancer. Its construction is based on a novel idea of PET tomography based on plastic scintillators and wavelength shifter~(WLS) and a new concept of positronium imaging.
We have prepared a simulation program based on Monte Carlo methods for optimizing the geometry and material of the J-PEM prototype.
Here we present the first results from the simulations and a brief review of the state of art of breast imaging modalities and their characteristics motivating our investigation.
\end{abstract}
\PACS{PACS numbers come here}
  
\section{Introduction}
Breast cancer is the most common cancer and the leading cause of death in females.
A woman's risk of developing invasive breast cancer in her life is approximately 1 in 8 (12\%)~\cite{WeidnerN}.
Diagnosis of breast cancer at earlier stages can result in improved outcomes when followed by timely and appropriate treatment, which allows for simpler and more cost-effective treatment to reduce both morbidity and mortality. 
The main point of making a diagnostic test is to decrease doctors doubt that a patient has a specific sickness, to the degree that they can settle on the board choices~\cite{JongRA}.
There is basic estimation of interpretation performance of screening tests which are based on sensitivity, specificity, positive predictive value, and cancer stage at diagnosis~\cite{Robert}. 
Sensitivity and specificity describe what a test can and cannot tell us, respectively. 
Both are expected to completely comprehend a tests qualities as well as its weaknesses. 

In more detail, sensitivity is expressed in percentage and defines the proportion of true positive subjects with the disease in a total group of subjects with the disease (True Positive/(True Positive+False Negative)).
Sensitivity is defined as the probability of getting a positive test result in subjects with the disease.
Hence, it relates to the potential of a test to recognise subjects with the disease~\cite{Rajul}.

Specificity is the number of true negative cases divided by the sum of true negatives and false positives.
In other words, specificity represents the probability of a negative test result in a subject without the disease.
Therefore, we can postulate that specificity relates to the aspect of diagnostic accuracy that describes the test ability to recognise subjects without the disease, i.e. to exclude the condition of interest.

The positive predictive value is the likelihood of having invasive cancer if one is recalled for assessment (true positive cases divided by the sum of the true positive and false positive cases).
A test that is highly sensitive will flag almost everyone who has the disease and not generate many false-negative results and a high-specificity test will effectively preclude nearly everyone who does not have the disease and will not create many false-positive results~\cite{Rajul}.
Sensitivity and specificity is rather small for all the conventional methods used for detection of breast cancer which leads to the problem that, for women  between ages 50 and 69 who go for the breast check-up, around 10\% of women are ill but not treated. Between 20 and 25\% of women, depending on the detection method, are wrongly diagnosed to have a cancer which leads to unnecessary biopsies~\cite{AM,sung}.
Because of the small dimensions of these lesions, a few millimeters in diameter, a detector system for such application must have high sensitivity and good spatial resolution.
Nowadays in present imaging modalities, sensitivity is compromised in the case of dense breasts, where it is difficult to distinguish between the tumor and the normal tissues.
Therefore, new diagnosis processes and systems for breast cancer are the subject of much research effort. One such research line relies on the use of Positron Emission based technology.
This is the case of the development of the J-PEM scanner, based on positron emission mammography (PEM) system, which aims at the detection of tumors with diameters down to 1 mm.
The requirements for high sensitivity and good spatial resolution can be fulfilled by J-PEM which combines wavelength shifters along with plastic scintillators, compact photo-detectors, large angular acceptance, depth-of-interaction (DOI) measurement capability and efficient data acquisition systems.

In this paper the J-PEM prototype for positron emission mammography is presented.
Focus is given to the system design and preliminary results from simulations for the validation of the detection techniques and system performance.
In Section 2 we have described possible imaging modalities with the given sensitivity and specificity.
Nevertheless, neither sensitivity nor specificity are influenced by the disease prevalence, meaning that results from one study could easily be transferred to some other setting with a different prevalence of the disease in the population.
Nonetheless, sensitivity and specificity can vary greatly depending on the spectrum of the disease in the studied group~\cite{Ana}.
Section 3 details the concept behind the J-PEM and the possibilities of positronium imaging~\cite{moskal,daria}.
Section 4 presents preliminary results using Monte carlo simulation, performed using the GATE toolkit. 
 
 \section{Screening test}
 The goal of screening asymptomatic women is to find breast cancer in its earliest stages when treatment has the highest chance for survival.
 For diagnosis and characterization of primary breast lesions, anatomical imaging such as mammography is the most common screening test for breast cancer,
 which is basically an x-ray picture of the breast. Examples of mammography images are shown in Fig. 1 for a 67 year old patient.
 Mammography may find tumors that are too small to feel.
 It may also find ductal carcinoma in situ (DCIS)~\cite{VirginiaL}.
 Mammography is more averse to discover breast tumors in women with dense breast tissue.
 Since both tumors and dense breast tissue seems white on a mammogram, it is hard to find a tumor when there is a dense breast tissue.
 The mammography sensitivity varies in the range depending on the age of the examined group and number.
 This range for sensitivity varies from 80\% to 96\% and in case of specificity it is in the range of 15 to 51.8\%~\cite{Luczynska}.
\begin{figure}[h]
\includegraphics[width = 0.43\textwidth]{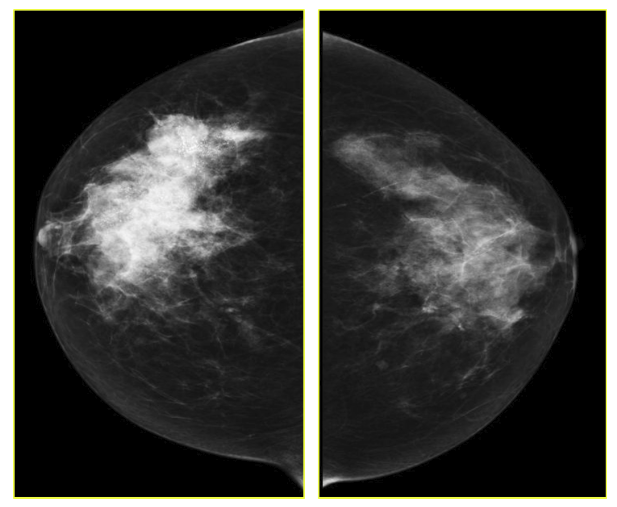}
\includegraphics[width = 0.4\textwidth]{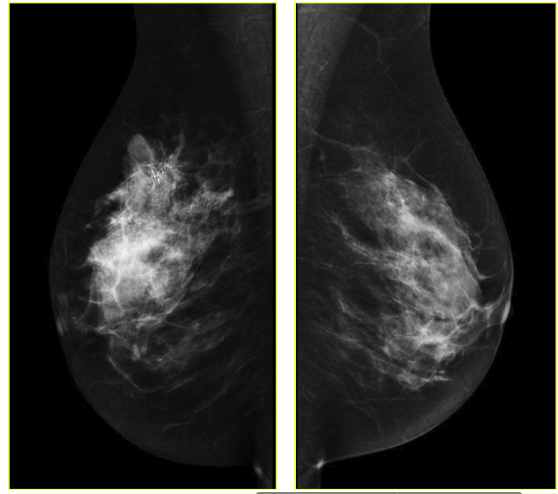}
\caption{Example of mammography images. Patient 67y/o. Left side – CC (cranio-caudal) views; Right side – MLO (mediolateral) views of both breasts. In the right breast, upper outer quadrant, approx. 6 cm from the nipple cluster of microcalcifications is visible (50x70mm). In the background of these microcalcifications well circumscribed, dense area (diameter 20mm) is present. BI-RADS cat.V. This image is provided by  Maria Sklodowska-Curie Memorial Institute of Oncology in Cracow, Poland.}
\end{figure}

However there is also a novel technique developed in the last few years based on the mammography methodology: contrast-enhanced spectral mammography (CESM). This method, like MRI, is based on imaging of tumor neoangiogenesis by use of a contrast agent~\cite{WeidnerN,DaniauxM}.
CESM uses a chelated iodine-base x-ray contrast agent, while MRI uses a chelated gadolinium-based paramagnetic agent ~\cite{LeachMO,BadrS}.
Because of high sensitivity of CESM (similar to sensitivity of MRI) this technique maybe comparable with MRI in some cases.
CESM, like other diagnostic methods, has some limitations.
Benign lesions enhance on CESM, just as they do on MRI.
That is, there is no possibility to generate a time-enhancement curve, comparable to that in breast MRI~\cite{Luczynska,E}.
Just as with mammography, CSEM sensitivity and specificity also varies depending on the age of the examined group and number. 
Sensitivity and specificity varies from 92.7\% to 100\% and 41\% to 69.7\%, respectively~\cite{Luczynska}. Exemplary CSEM images for CC and MLO views are shown in Fig. 2.
\begin{figure}[h]
\includegraphics[width = 0.42\textwidth]{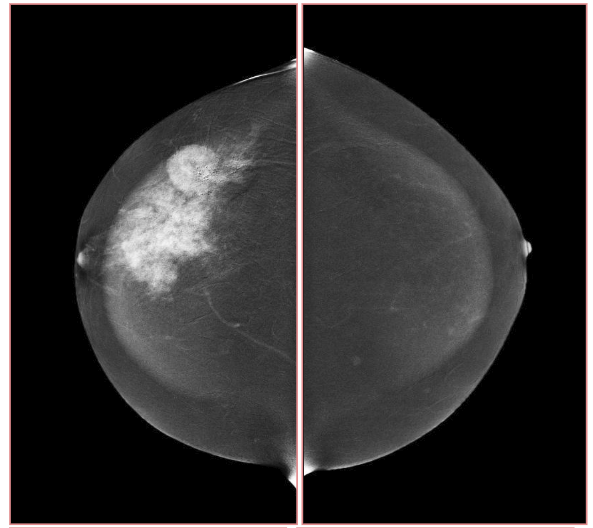}
\includegraphics[width = 0.4\textwidth]{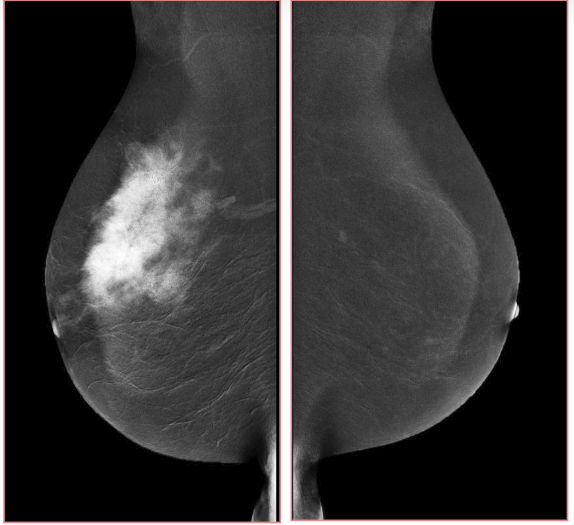}
\caption{Example of CESM images of the same 67 y/o patient as on Fig 1, left side – CC (cranio-caudal) views; right side – MLO (mediolateral) views of both breasts. Right breast, upper outer quadrant, pathological contrast enhancement area is visible (dimensions 110x75 mm). BI-RADS category V. This image was provided by  Maria Sklodowska-Curie Memorial Institute of Oncology in Cracow, Poland.}
\end{figure}

Digital breast tomosynthesis (DBT) is a promising new technology for acquiring and displaying three-dimensional mammograms~\cite{Per}.
Because of its improved ability to differentiate true breast lesion or summation of normal breast structures.
An example image of tomosynthesis is shown in Fig. 3, for the different planes, it has improved characterization of masses. Tomosynthesis is increasingly being used in the diagnostic setting to evaluate masses, asymmetries, and architectural distortion.
Tomosynthesis can be done in full or spot compression views as needed~\cite{LTNiklason}.
The sensitivity and specificity of DBT ranged from 74.2\% to 86.9\% and 97.0\% to 97.5\%, respectively~\cite{LeiJ}.
\begin{figure}[h]
\includegraphics[width = 0.8\textwidth]{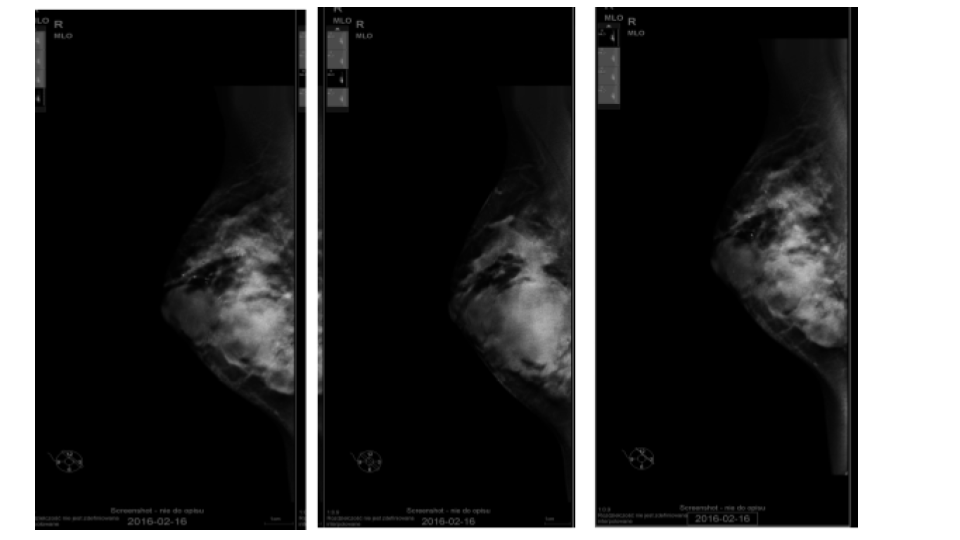}
\centering
\caption{ Example of Tomosynthesis planes of 29 y/o patient - in the planes there are numerous well limited shadows architectural distortion and microcalcifications on histopathology - papillary carcinoma in the right breast.
Breast thickening observed for 3 years. The right breast ultrasound, visualized numerous solid and fluid lesions.
This image was provided by  Maria Sklodowska-Curie Memorial Institute of Oncology in Cracow, Poland.}
\end{figure}

Ultrasound is well accepted as the most useful adjunct to mammography for the diagnosis of breast abnormalities. 
An exemplary ultrasound image of right breast of a patient is shown in Fig. 4.
Ultrasound is most often used to assess palpable masses and nonpalpable masses that have been detected during screening mammography~\cite{Durfee,Frazier}.
Ultrasound may demonstrate malignancies and other masses that are not visible mammographically~\cite{Gordon}. 
Ultrasound had an overall pooled sensitivity and specificity of 80.1\% and 88.4\%, respectively~\cite{Luczynska}.
\begin{figure}[h]
\includegraphics[width = 0.6\textwidth]{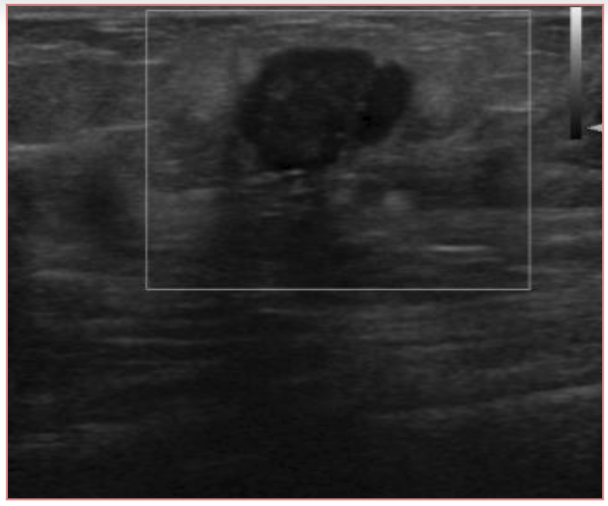}
\centering
\caption{Example image of breast ultrasound (the same patient as on Fig.1). Right breast, outer upper quadrant, 7cm from the nipple not well defined hypoechogenic area (dimensions: 24x20mm) is visible. BI-RADS category IV c.
This image was provided by  Maria Sklodowska-Curie Memorial Institute of Oncology in Cracow, Poland.}
\end{figure}

Molecular breast imaging (MBI) is a nuclear medicine technique that uses dedicated gamma cameras to image the physiologic uptake of a radiopharmaceutical, typically 99mTc-sestamibi, in the breast.
MBI is capable of detecting mammographically occult cancers, particularly in women with dense breasts~\cite{Rhodes,Rhodes1}.
MBI has an overall sensitivity of 90\%, with a sensitivity of 82\% for lesions less than 10 mm in size~\cite{connor}. 

Magnetic resonance imaging (MRI) has been used as an adjunctive screening tool, mainly for women who may be at increased risk for the development of breast cancer.
Example images of MRI are shown in fig. 5 with T1 contrast enhancement.
MRI for screening has not been very popular in women with average risk due to concerns about the low specificity leading to additional biopsies, time and cost of technology~\cite{selvi}.
Breast MRI sensitivity values reported in high risk screening studies range from 93\% to 99\%.
Despite its high sensitivity, breast MRI has been reported to have variable specificity, ranging from 50\% to 85\%~\cite{med}. 
These number of sensitivity and specificity depends on the type of tumor, size of tumor, age of patient, and where it is localised. 
\begin{figure}[h]
\includegraphics[width = 0.4\textwidth]{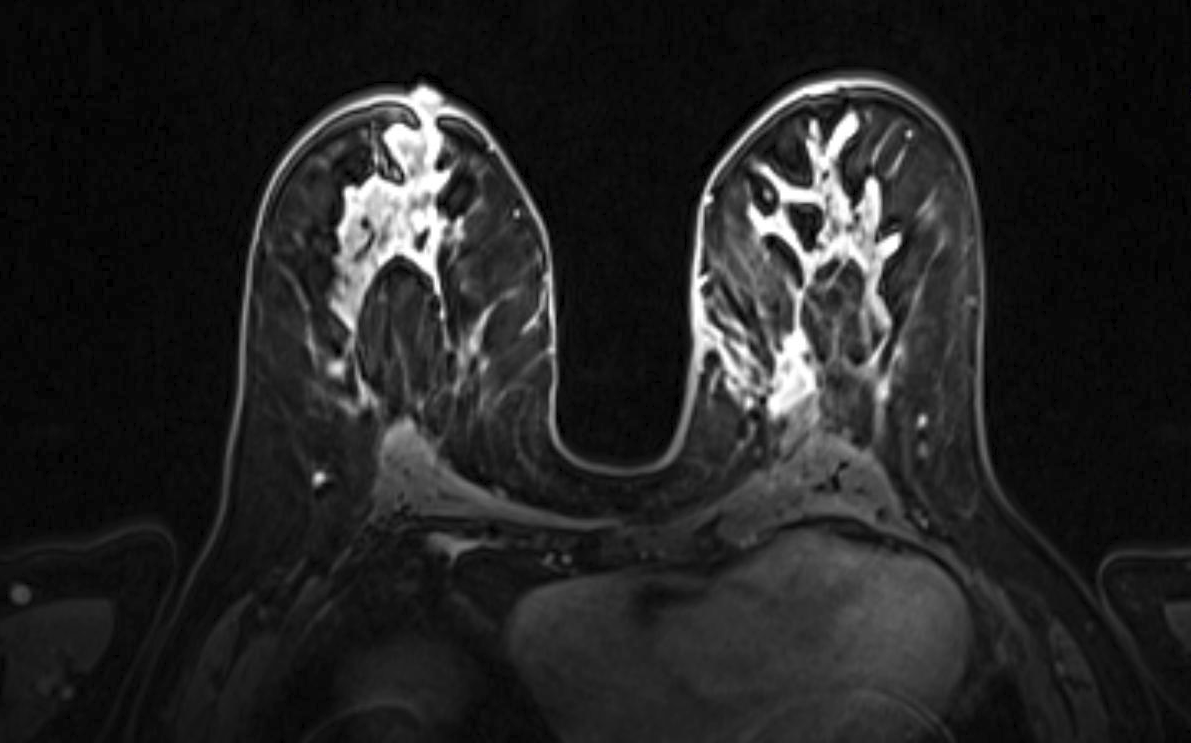}
\includegraphics[width = 0.4\textwidth]{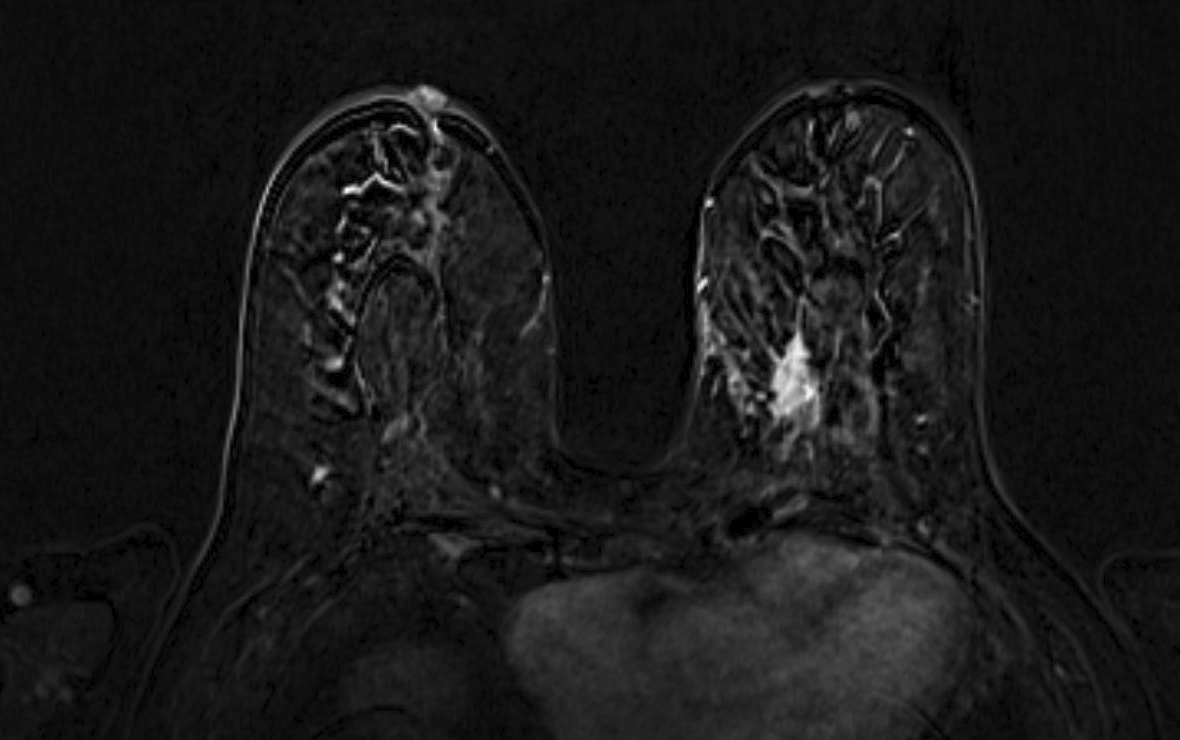}
\includegraphics[width = 0.4\textwidth]{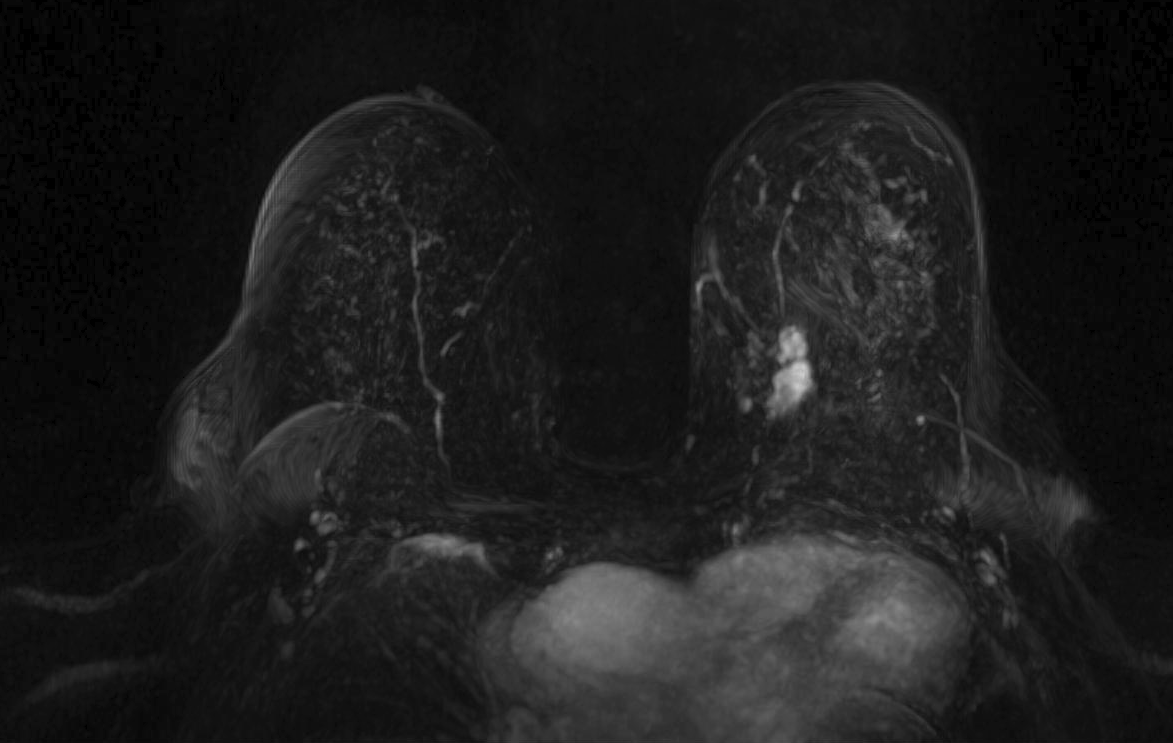}
\centering
\caption{ Example images of breast MRI (MIP, subtraction T1 contrast enhanced image, T1 contrast image with fat saturation).
This image is provided by  Maria Sklodowska-Curie Memorial Institute of Oncology in Cracow, Poland.}
\end{figure}

Then there is positron emission mammography which is a new technology that is designed for the imaging of specific small body parts.
To achieve high resolution it uses short-lived positron-emitting isotopes to generate images of cancer within the breast~\cite{shannon}.
It has better specificity as compared to MRI, which confines the unneeded biopsies. The potential basis for the improved sensitivity of PEM is enhanced spatial resolution~\cite{JS}, which enables detection of small lesions.
This is our motivation for developing J-PEM.
PEM has a higher imaging sensitivity in small tumors $<2$ cm. Even in very small tumors $< 1$ cm, the imaging sensitivity was acceptable at 60-70\%~\cite{JS}.
The standard sensitivity and specificity of PEM is 80\% and 100\%, respectively~\cite{kavita}.

\section{Concept of J-PEM}
The J-PEM is a prototype intended to evaluate PET technology in the diagnosis of malign neoplasm in the breast and of ganglion loco-regional invasion.
It is based on plastic scintillators and utilizes the same technology as the 
Jagiellonian Positron
Emmision Tomograph,
J-PET~\cite{shivani}.
It is optimized for the detection of photons from electron-positron annihilation~\cite{P,PM}.
Such photons, having an energy of 511 keV, interact with electrons in plastic scintillators predominantly via the Compton effect.
J-PEM uses a dedicated instrument for breast cancer detection that is equipped with two parallel photon detectors in a configuration similar to mammography compressors.
The detector system consists of two modules of plastic scintillators, with each module built from two layers of plastic scintillator and the wavelength shifters~\cite{J,smy} placed orthogonally between them, as shown in Fig. 6.
Each scintillator bar is attached at both ends with Silicon Photomultipliers for the signal readout~\cite{P}.
The combined use of plastic scintillators, which have superior timing properties, with the WLS strips can provide an affordable and precise scanner with significant improvement in spatial resolution and efficiency for the detection of breast cancer.
Plastic scintillators are characterized by short light decay time which is in the order of 1.5 ns~\cite{P}. This enables one to achieve high time resolution.
In the J-PET, solution for the position of the interaction point of photons is based on the measurement of the time of the signals arrival to the ends of the long scintillator strips.
So far a resolution of about 10 mm was achieved~\cite{smy}.
In order to achieve high resolution, we propose to register scintillation light escaping the scintillator bar through a side wall using an array of WLS.
It has been already proven that one can reach to position resolution of 5 mm for the coordinate along the scintillator bar.
\begin{figure}[h!]\centering
\includegraphics[width =0.7\textwidth]{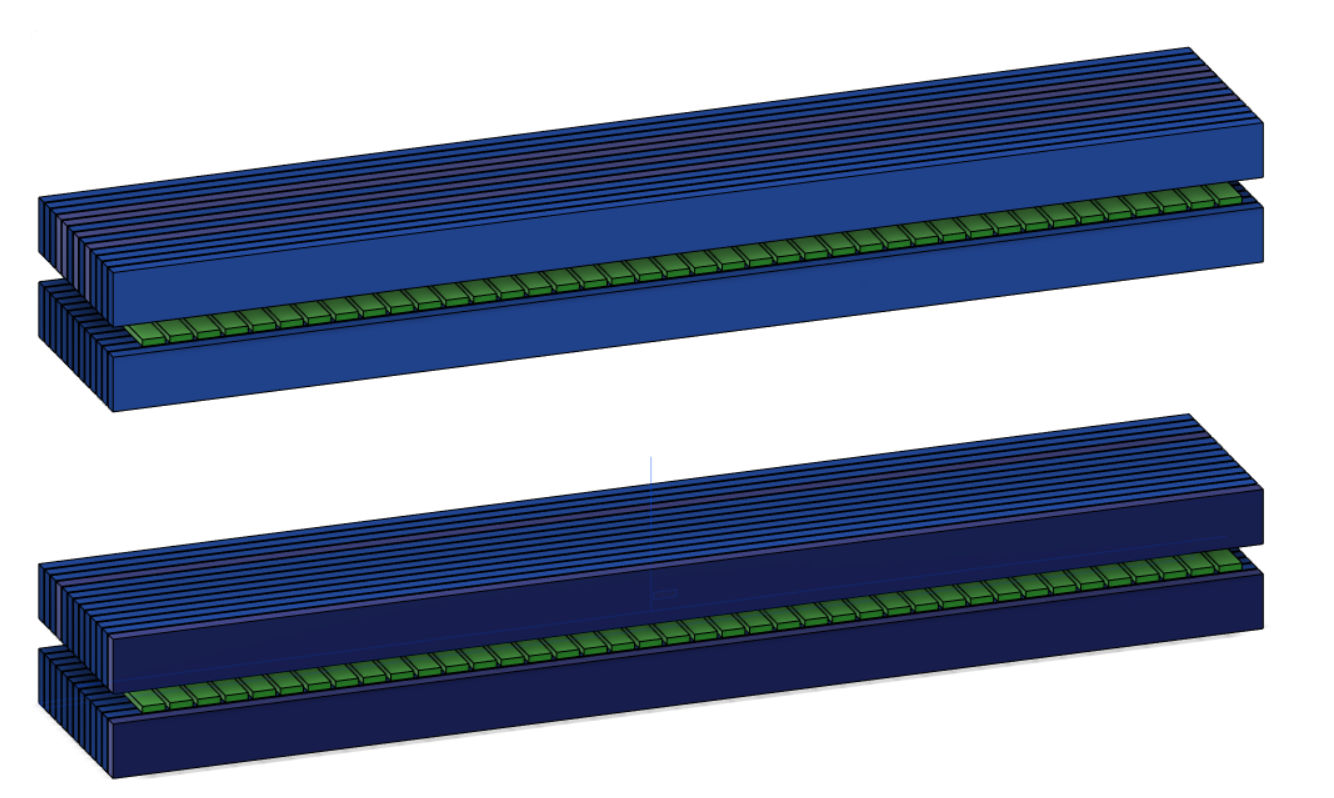}
\includegraphics[width=5cm,height=4cm]{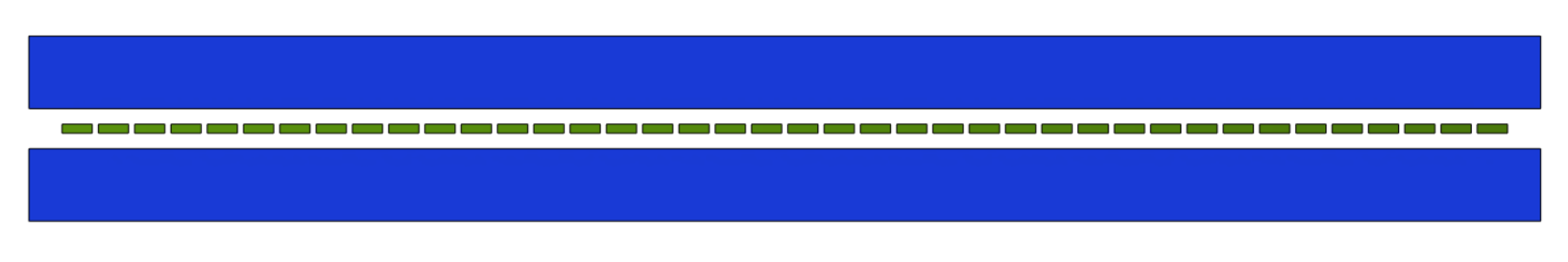}
\includegraphics[width =0.5\textwidth]{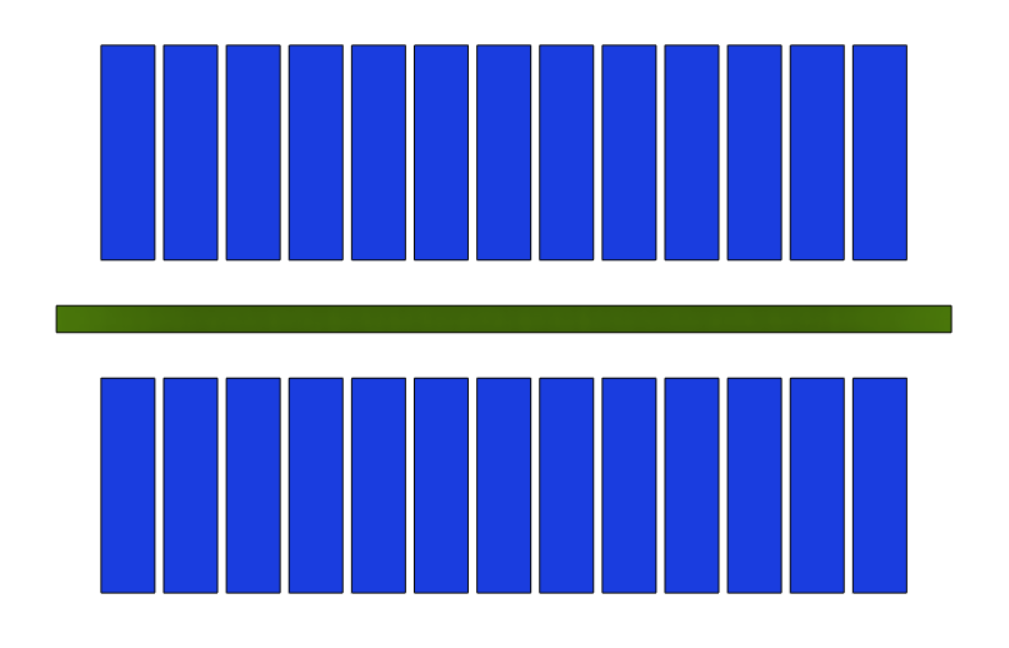}
\caption{Top figure: Geometry scheme of J-PEM used in GATE simulation. Here blue color represents plastic scintillators and green color represents wavelength shifters. Dimension of plastic scintillators and WLS were 6x24x500 mm and 3x10x100 mm, respectively. Space between the modules is 33.88 cm. Bottom Left and right are the Y-Z and X-Y plane view of the given geometry for single module.}
\end{figure}
J-PET is capable of performing simultaneous imaging of the density distribution of annihilation points as well as positron annihilation lifetime spectroscopy~\cite{moskal,Dulski}.
The positronium imaging technique has successfully distinguished cancer from healthy tissue by in vitro probing of human tissues or in the in vitro organoid (3D cell culture) systems~\cite{Axpe}.
Study of positronium decay can provide  new data in medical diagnosis as the mean lifetime of positronium depends on the size of free volume between atoms, whereas its formation probability depends on the concentration of voids~\cite{zuza}.
In the free space between atoms, positronium decays as it does in vacuum~\cite{moskal}.
However, within molecules there are additional annihilation possibilities and the mean lifetime of ortho-positronium decreases significantly compared with the lifetime in vacuum, from 142 ns to a few nanoseconds~\cite{B}.
The reconstruction of positronium lifetime requires determination of times of its creation and annihilation. These can be achieved when applying radiopharmaceuticals labeled with isotopes such as
Scandium-44 which after emission of the positron changes into a daughter nucleus in an excited state ~\cite{daria}.
A technique that combines PALS and PET in clinical use must enable determination of positronium parameters in a position-sensitive manner, and it needs to be scaled to work for living organisms.
Recently, a first possible solution designed for the size of a human body was proposed.
In vitro studies comparing the positronium properties in cancerous and healthy tissues suggest that the ortho-positronium mean lifetime is correlated with the grade  of development of metabolic disorders in cancer cells~\cite{zuza,B}.
Combining the information about metabolism with information about the structure of tissue will improve specificity and sensitivity, therefore increasing the accuracy of the examination.

\section{Monte Carlo simulation}
In order to quantify the J-PEM geometry we have performed Monte-Carlo simulations. The simulations were performed using the GATE package.
GATE (Geant4 Application for Tomographic Emission) is a Monte Carlo simulation platform developed by the Open-GATE collaboration~\cite{gate1} based on Geant4 software.
It is dedicated to numerical simulations in medical imaging and radiotherapy. It utilizes an easy macro mechanism to configure the experimental settings for computed tomography, single photon emission computed tomography, positron emission tomography as well as optical imaging (bioluminescence and fluorescence) or radiotherapy.
In the simulations the full geometry of the J-PEM detector and the composition of the detector material were taken into account. 
The interactions of photons in the scintillators were simulated by GATE.
In the simulations we assumed that annihilation source is placed in the center of the detector and that the back-to-back photons (each with energy of 511 keV) from the Ps$\rightarrow2\gamma$ annihilation are isotropically emitted.
Energy deposition inside the plastic scintillator is shown in Fig.7. According to the performed simulation we are able to register 4.33\% of generated back to back events.
After adding the condition that both photons deposited at least 200 keV energy fraction of register annihilation drop to  0.8 \% .
Such a condition is needed for the suppression of the scatter fraction ~\cite{kow}.
We found that there was 11.3\% primary back-to-back registered events where at least one photon scattered in WLS.
\begin{figure}[h!]
\includegraphics[width = 0.45\textwidth]{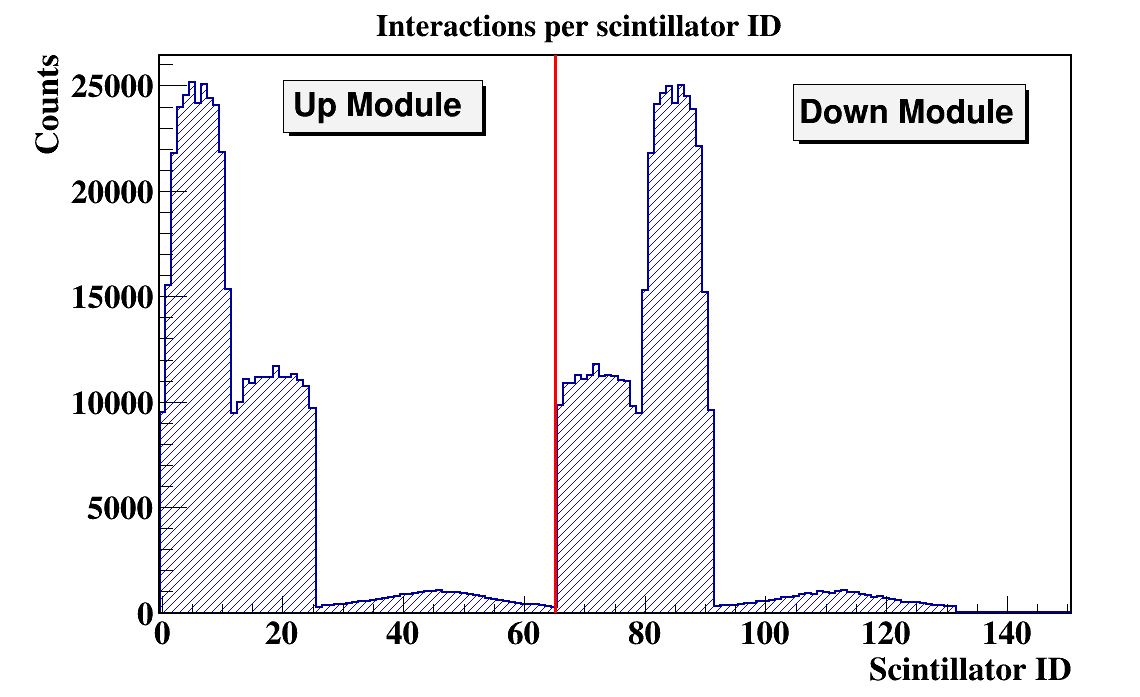}
\includegraphics[width = 0.45\textwidth]{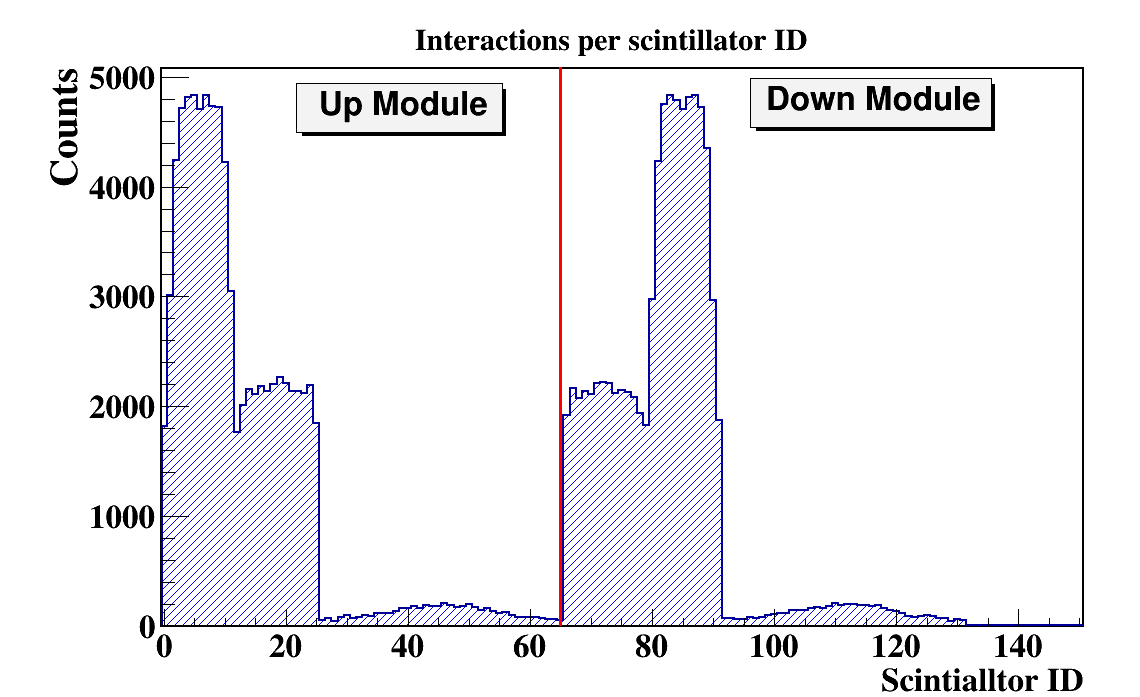}
\includegraphics[width = 0.45\textwidth]{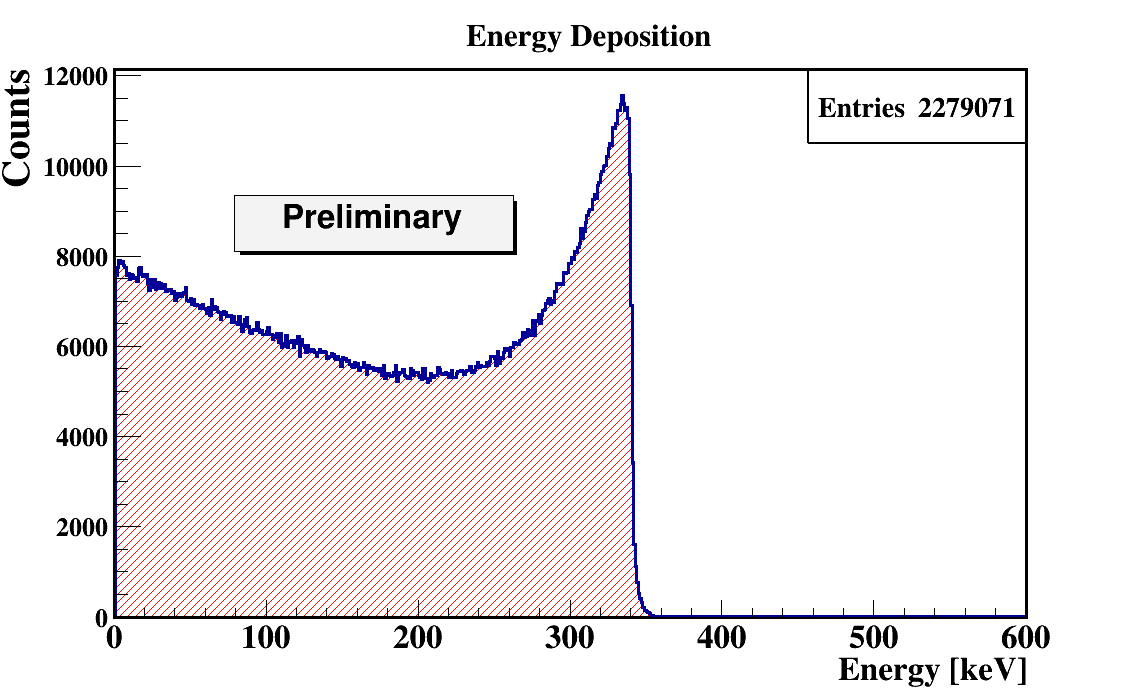}
\centering
\caption{Top left : Shows the number of interactions per scintillator ID for back to back photons.Top right: Shows the number of interactions per scintillator ID for back to back photons with condition that both photons deposited at least 200 keV energy.
Scintillators in the inner and outer layer of upper module are numbered from 1 to 13 and 14 to 26, respectively.
While, WLS in the upper module are numbered from 27 to 67.
Scintillators in the inner and outer layer of down module are numbered from 68 to 81 and 82 to 94, respectively.
WLS numbering for down module are from 95 to 135.
Figure in bottom shows energy deposition of primary photons.}
\end{figure}
The J-PEM simulation will be used to evaluate the performance of various detector configurations and options. Sensitivity estimations, study of the depth-of-interaction measurement technique and estimation of the background rates are some examples of basic information to be obtained from the simulation and that is needed to finalize the detector concept.
Finally, simulated events will also be used as input to the reconstruction software.

\section{Conclusion}

In this paper we have investigated the design, construction and establishment of the characteristic performance of the Jagiellonian Positron Emission Mammography (J-PEM) for the detection and diagnosis of breast cancer. Its construction is based on a novel idea of PET based on plastic scintillators and wavelength shifter~(WLS) and a new concept of positronium imaging. We have prepared a simulation program based on Monte Carlo method using GATE toolkit, for optimizing the geometry and material for the J-PEM prototype. We have also presented the first results from the simulations and a brief review of the state of art of breast imaging modalities and their characteristics motivating our investigation. Results from the simulations are showing that the amount of registered back to back events are comparable to those of conventional secondary screening methods for breast cancer~\cite{David},
thus making J-PEM a viable tool in early detection of cancer.

\end{document}